\documentclass{aa}  
\usepackage{graphicx}
\usepackage{color}
\usepackage{hyperref}
\usepackage{longtable,ltcaption}
\usepackage{lscape}
\usepackage{natbib}
\usepackage{lscape}
\usepackage{amsmath}
\usepackage[nointegrals]{wasysym}
\usepackage{graphics}
\usepackage{times}
\usepackage{parskip}
\usepackage{pdflscape}
\usepackage{geometry}
\usepackage{marginnote}
\usepackage{multicol}
\usepackage{soul}
\usepackage{amsfonts}
\usepackage{txfonts}

\newfont{\nlx}{cmssdc10 scaled 900}
\newfont{\mfont}{cmssdc10 scaled 810}
\definecolor{mygreen1}{rgb}{0.0,0.604,0.131} 
\definecolor{myblue1}{rgb}{0.0,0.604,0.831} 
\definecolor{myblue2}{rgb}{0.0,0.49,0.6745}
\definecolor{myblue3}{rgb}{0.0156,0.4078,0.9921}
\definecolor{myblue4}{rgb}{0.0,0.44,0.87}
\definecolor{myred1}{rgb}{0.529,0.019,0.017}
\definecolor{mycyan}{rgb}{0.63921569,0.0,0.48235294}

\newcommand{\brem}[1]{\textcolor{black}{\nlx #1}}
\newcommand{\mbrem}[1]{\textcolor{black}{\mfont #1}}

\hypersetup{
    pdftoolbar=true,				
    pdfmenubar=true,				
    pdffitwindow=false,			
    pdfstartview={FitH},			
    pdftitle={title},				
    pdfauthor={Author},			
    pdfsubject={Subject},			
    pdfcreator={Author},			
    pdfproducer={Author},			
    pdfkeywords={Population \& Evolutionary} ,        
    pdfnewwindow=true,			
    colorlinks=true,				
    linkcolor=cyan,				
    citecolor=myblue4,				
    filecolor=cyan,				
    urlcolor=cyan				
}

\def\SFQ{\mbrem{SFQ}}
\def\ioSFQ{\mbrem{ioSFQ}}
\def\tsfq{$\tau_{\rm SFQ}$}
\def\msun{$\mathrm{M}_{\odot}$}

\def\starlight{{\sc Starlight}}

\def\D4000{$D_{4000}$}

\def\rr{$\mathrm{R}^{\star}$}
\def\rbulge{$\mathrm{R}_{\rm B}$}
\newcommand{\sbb}{mag/$\sq\arcsec$}
\newcommand{\dmb}{$<\!\!\!\delta\mu_{9{\rm G}}\!\!\!>$}

\def\pcrit{log$\mathrm{(}M,\Sigma\mathrm{)}_{\star}^{\rm c}$}
\def\acrit{log$\mathrm{(}M,\Sigma\mathrm{)}_{\star,{\rm B}}$}
\def\ms{$\mathrm{(}M,\Sigma\mathrm{)}_{\star}$}

\def\mbstar{${\cal M}_{\star,\textrm{B}}$}
\def\mbulge{${\cal M}_{\star,\textrm{B}}$}
\def\mstotal{${\cal M}_{\star,\textrm{T}}$}

\def\tottlight{$\langle t_{\star} \rangle_{{\cal L}}$}
\def\ytottlight{$\langle t_{\star}/{\rm yr} \rangle_{{\cal L}}$}

\def\mtmass{$\langle t_{\star,\textrm{B}} \rangle_{{\cal M}}$}

\def\stmass{$t_{\star,{\cal M}}$}
\def\stlight{$t_{\star,{\cal L}}$}

\def\tsstar{$\Sigma_{\star}$}
\def\sstar{$\Sigma_{\star,\mathrm{B}}$}

\newfont{\hvss}{cmssdc10 scaled 1540}
\def\?{{\bf\color{red}?}}
\def\reff{$\mathrm{R}_{\rm eff}$}
\def\ha{H$\alpha$}

\def\tmass{$\langle t_\star \rangle_{{\cal M}}$}
\def\tlum{$\langle t_\star \rangle_{{\cal L}}$}

\def\kmsec{km\,s$^{-1}$}
\def\ewha{EW(H$\alpha$)}
\def\lyc{Ly{\sc c}}

\newcommand{\SL}{{\sc Starlight}}
\def\rb{$\mathrm{R}_{\rm B}$}
\def\gradtt{$\mathrm \nabla ( t_{\star,B})_{\cal L,M}$}
\def\gradtM{$\mathrm \nabla ( t_{\star,B})_{\cal M}$}
\def\gradtL{$\mathrm \nabla ( t_{\star,B})_{\cal L}$}

\def\sisp{\nlx sisp\rm}
\def\isan{\nlx isan\rm}

\def\vq{$v_{\rm q}$}
\def\mmu{$\langle\mu\rangle$}

\begin{document} 

\title{Stellar age gradients and inside-out star formation quenching in galaxy bulges}
   \author{
          Iris Breda
         \inst{\ref{IA-CAUP},\ref{UPorto}}          
          \and
          Polychronis Papaderos
          \inst{\ref{IA-CAUP},\ref{IA-FCiencias},\ref{ULisbon}}
          \and
          Jean Michel Gomes
          \inst{\ref{IA-CAUP}}          
          \and 
          Jos\'e Manuel V\'ilchez
          \inst{\ref{IAA-CSIC}}
          \and 
          Bodo L. Ziegler
          \inst{\ref{UWien}}
          \and          
          Michaela Hirschmann
          \inst{\ref{DARK}}
          \and          
          Leandro S.M. Cardoso
          \inst{\ref{IA-CAUP},\ref{UEvora}}
          \and 
          Patricio Lagos
          \inst{\ref{IA-CAUP}}
          \and
          Fernando Buitrago
          \inst{\ref{IA-FCiencias},\ref{ULisbon}}          
          }
\institute{Instituto de Astrof\'{i}sica e Ci\^{e}ncias do Espaço - Centro de Astrof\'isica da Universidade do Porto, Rua das Estrelas, 4150-762 Porto, Portugal \label{IA-CAUP}
         \and
Departamento de F\'isica e Astronomia, Faculdade de Ci\^encias, Universidade do Porto, Rua do Campo Alegre, 4169-007 Porto, Portugal \label{UPorto}
         \and
Instituto de Astrofísica e Ciências do Espaço, Universidade de Lisboa, OAL, Tapada da Ajuda, PT1349-018 Lisboa, Portugal 
\label{IA-FCiencias}
         \and 
Departamento de Física, Faculdade de Ciências da Universidade de Lisboa, Edifício C8, Campo Grande, PT1749-016 Lisboa, Portugal \label{ULisbon}    
\and 
DARK, Niels Bohr Institute, University of Copenhagen, Lyngbyvej 2, 2100 Copenhagen, Denmark
\label{DARK}
\and 
 Instituto de Astrof\'isica de Andaluc\'ia (CSIC), Glorieta de la Astronom\'ia s/n, 18008 Granada, Spain
 \label{IAA-CSIC}
\and 
 Department of Astrophysics, University of Vienna, T\"urkenschanzstr. 17, A-1180 Vienna, Austria
 \label{UWien}
\and 
Department of Mathematics, University of \'Evora, R. Romao Ramalho 59, 7000 \'Evora, Portugal
\label{UEvora}
\\
             \email{iris.breda@astro.up.pt}
             }

\date{Received ??; accepted ??}
\abstract{Radial age gradients hold the cumulative record of the multitude of physical processes driving the build-up of stellar populations and the ensuing star formation (SF) quenching process in galaxy bulges, therefore potentially sensitive discriminators between competing theoretical concepts on bulge formation and evolution. 
Based on spectral modeling of integral field spectroscopy (IFS) data from the CALIFA survey, we derive mass- and light-weighted stellar age gradients (\gradtt) within the photometrically determined bulge radius (\rbulge) of a representative sample of local face-on late-type galaxies that span 2.6 dex in stellar mass (8.9 $\leq$ log\mstotal $\leq$ 11.5). Our analysis documents a trend for decreasing \gradtt\ with increasing \mstotal, with high-mass bulges predominantly showing negative age gradients and vice versa. The inversion from positive to negative \gradtt\ occurs at log\mstotal $\simeq$ 10, which roughly coincides with the transition from lower-mass bulges whose gas excitation is powered by SF to bulges classified as Composite, LINER or Seyfert.  
We discuss two simple limiting cases for the origin of radial age gradients in massive LTG bulges. The first one assumes 
that the stellar age in the bulge is initially spatially uniform (\gradtt\ $\approx$ 0), thus the observed age gradients 
($\sim$ --3 Gyr/\rbulge) 
arise from an inside-out SF quenching (\ioSFQ) front that is radially expanding with a mean velocity \vq. 
In this case, the age gradients for massive bulges translate into a slow (\vq\ $\sim$1-2 \kmsec) \ioSFQ\ that lasts until $z \sim2$, 
suggesting mild negative feedback by SF or an AGN.
If, on the other hand, negative age gradients in massive bulges are not due to \ioSFQ\ but primarily due to their inside-out formation process,
then the standard hypothesis of quasi-monolithic bulge formation has to be discarded in favor of a scenario that involves gradual buildup of stellar mass over 2-3 Gyr through, e.g., inside-out SF and inward migration of SF clumps from the disk.
In this case, rapid ($\ll$ 1 Gyr) AGN-driven \ioSFQ\ cannot be ruled out.
While the \mstotal\ vs. \gradtt\ relation suggests that the assembly history of bulges is primarily regulated by galaxy mass,
its large scatter ($\sim$1.7 Gyr/\rbulge) reflects a considerable diversity that calls for an in-depth examination of the role of various processes (e.g., negative and positive AGN feedback, bar-driven gas inflows) with higher-quality IFS data in conjunction with advanced spectral modeling codes.
}
\maketitle

\parskip = \baselineskip

\section{Introduction \label{intro}}
A key element in our understanding of bulge evolution in late-type galaxies (LTGs) revolves around the termination of their dominant mass assembly phase through star formation quenching (\SFQ). This process is thought to result from different, perhaps non-mutually exclusive mechanisms, such as morphological quenching \citep[stabilization of the disk against gas fragmentation once its 
center becomes dominated by a massive stellar spheroid,][]{Martig09,Genzel14}, inhibition of inflow of cold gas from the cosmic web 
due to virial shocks in the galactic halo \citep{Dekel09} or stripping away of the gaseous reservoir of galaxies in clusters
\citep[][see also, \textcolor{myblue4}{Peng et al. 2015}]{LarsonTinsleyCaldwell1980}, or negative feedback by an active galactic nucleus \citep[AGN; e.g.,][]{Silk97,DMSH05,Croton06,Cattaneo09}. 

Various lines of evidence suggest that \SFQ\ is initiated once galaxy bulges have grown to a mass \mbstar\ $\sim 3 \times 10^{10}$ \msun\ \citep{Strateva01} and a stellar surface density \sstar\ $\sim 10^9$ \msun\,kpc$^{-2}$ \citep{Kauffmann03,Gon16}.  
Indeed, whereas in the local universe star formation (SF) is almost omnipresent in low-\mbstar, low-\sstar\ bulges, 
it is steeply vanishing above a characteristic mass-density threshold \pcrit\ $\simeq$ (10,9), as documented through multi-band photometry \citep[e.g.,][]{Peng10,Omand14}, single-aperture spectroscopy \citep[e.g.,][]{Kauffmann03,Bri04} and, more recently, spatially resolved integral field spectroscopy \citep[IFS; e.g.,][BP18]{Fang13,CatalanTorrecilla17,Ellison18,WE19}. 
For instance, \citet{Zibetti17} report from an analysis of stellar indices 
for 394 galaxies from the CALIFA IFS survey \citep{Sanchez12-DR1} a bi-variate distribution of galaxy stellar populations on the \tsstar\ vs. luminosity-weighted age \ytottlight\ plane, with old (log\tottlight\ $\simeq$ 10) quiescent E\&S0 galaxies populating a high-density peak (log(\tsstar) $\geq$ 8.3), whereas younger (log\tottlight\ $\leq$ 9.5) LTGs being confined to log(\tsstar) $\leq$ 8. 
\citet[][hereafter BP18]{BP18} find from spatially resolved modeling of CALIFA IFS data for a representative sample of local LTGs 
that the contribution \dmb\ of stellar populations younger than 9 Gyr to the bulge $r$-band mean surface brightness \mmu\ is tightly anti-correlated with \mbstar\ and \sstar, showing a monotonous decrease from $\sim$60\% in the lowest-mass bulges to $\la$ 10\% in the most massive and dense ones (\acrit\ $\simeq$ 11.3,9.7). 
This, and the fact that the bulge-to-disk age contrast increases with increasing \mstotal\ (BP18)
suggests that the assembly timescale of bulges scales inversely with galaxy mass, that is, the more massive a LTG is, the earlier it has experienced the dominant phase of its bulge build-up (what these authors termed sub-galactic downsizing), in agreement with earlier conclusions by \citet{Ganda07}.
This trend also adds further support to the picture of inside-out galaxy growth \citep{Eggen62,FE80,vdB98,Kepner99}, 
with the dense galaxy centers completing their assembly first while stellar mass continuing building up in the galaxy periphery, in agreement with previous findings \citep[e.g.,][]{MunozMateos07,Salim12,Perez13,Gon14b,Tacchella15}.
At a higher redshift $z$, the association between \pcrit\ and \SFQ\ is established through abundance matching studies \citep{vD13}, pixel-by-pixel SED fitting \citep{Wuyts12,Lang14,Tacchella15,Mosleh18} or empirical relations between rest-frame color and mass-to-light ratio \citep{Szomoru12}.

The growth of stellar spheroids above \pcrit\ and the ensuing \SFQ, appears to be accompanied by a gradual change in the dominant gas excitation mechanism: whereas nebular emission in low-mass bulges is typically powered by SF, the majority of bulges above \pcrit\ (for instance, 94\% of those bulges in the sample of BP18) fall in the loci of Seyferts, LINERs \citep[low-ionization nuclear emission-line regions;][]{Heckman80} and Composites. The affinity of quenched, high-\tsstar\ stellar spheroids (bulges and early-type galaxies-ETGs) to LINER-specific emission-line ratios \citep[e.g.,][see \textcolor{myblue4}{Kormendy \& Ho 2013} for a review]{Annibali10,YanBlanton2012}, recently documented with IFS out to several kpc from the nucleus \citep{P13,Singh13,G16a}, thereby adding justification to the generic term LIER \citep[e.g.][]{Belfiore16}, has been traditionally attributed to photoionization by old ($\geq$ 100 Myr) post-AGB sources  \citep{bin94} or a weak AGN \citep{Ho08}. \citet{P13}, on the other hand, argue that even a strong AGN cannot be ruled out in massive spheroids showing merely weak LINER emission (i.e. with an \ha\ equivalent width \ewha\ $<$ 0.5 \AA): this is because these authors \citep[see also][]{G16a} find that, in the absence of absorbing cold gas with a sufficient filling factor, the bulk of \lyc\ radiation from pAGB sources (consequently, also from a putative AGN) is escaping without being reprocessed into nebular emission. 
Indeed, the paucity of a cold medium in the nuclear region of quenched spheroids can be seen as natural consequence of depletion, thermalization and expulsion of gas by SF and eventually an AGN. 
If so, \SFQ, LyC photon escape, and LI(N)ER emission are actually inseparable and causally linked facets of one and the same phenomenon, namely the partial or complete evacuation of cold gas from bulges and ETGs once they have grown above \pcrit, and AGN, if present, has fully unfold its energy impact.
 
\begin{figure} 
\includegraphics[width=1\linewidth]{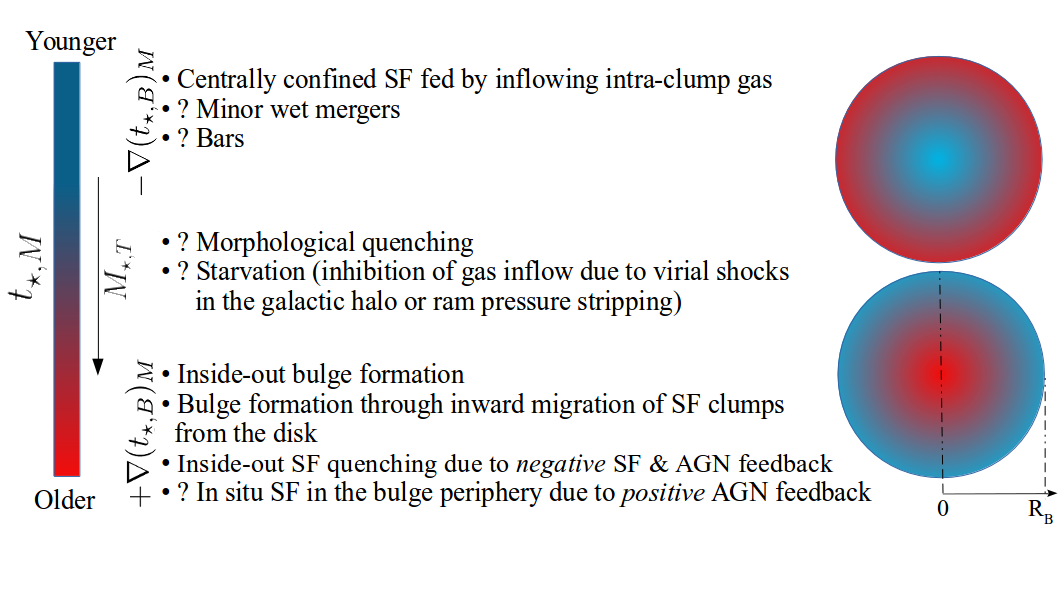}
\caption[]{Effect of different processes shaping the evolution of LTG bulges on the slope of the mass-weighted stellar age gradient \gradtM.}
\label{diagram}
\end{figure}

Whereas there is broad consensus that the dense galaxy centers form and quench first \citep[e.g.][see also \textcolor{myblue4}{Lin et al. 2019}]{Tacchella15,Tacchella17}, the timescale \tsfq\ of inside-out \SFQ\ (\ioSFQ) in LTG bulges is poorly constrained. 
A circumstantial argument that the onset of accretion-powered nuclear activity above \pcrit\ does not universally initiate a rapid \SFQ\ rests on the co-existence of bulges classified as Composite, LINER and Seyfert above \pcrit. 
This led BP18 to deduce a \tsfq\ $\ga$ 2 Gyr, in agreement with estimates by \citet{Tacchella15} from analysis of massive 
(10.84 $\leq$ log\mstotal\ $\leq$ 11.7) galaxies at $z$ $\sim2.2$.
Clearly, quantitative inferences on \tsfq\ as a function of total and bulge mass are desirable, given that they could help discriminating between different proposed \SFQ\ mechanisms. For instance, a short \tsfq\ 
(e.g., on the order of the warm-gas sound crossing time of 600 Myr for a typical LTG bulge with \rbulge$\sim$3 kpc)
would be consistent with a single or several intermittent energetic episodes leading to quick gas removal 
(e.g., most plausibly, a strong AGN outburst or a series of nuclear starbursts), and otherwise with a gradual inside-out 
depletion of gas or, possibly, morphological quenching, or gas starvation scenarios (cf. Fig.~\ref{diagram}).

Quick \ioSFQ\ should thus nearly preserve pre-existing stellar age gradients, whereas slow \ioSFQ\ produce or amplify negative age gradients. Therefore, in the idealized case of a bulge with initially uniform age that experiences an outwardly propagating \SFQ\ front with a constant radial velocity \vq, the slope (Gyr/kpc) in stellar age ($\propto 1/$\vq) could help placing   constraints on \vq\footnote{An analogous case of conversion of stellar age gradients into a mean SF propagation 
velocity was presented in \citet{P98} for the blue compact galaxy \object{SBS 0335-052E} \citep{Izotov90}.}.

Several observational studies in the past decades have greatly improved our understanding on age and star formation rate (SFR) patterns in galaxies and their bulges. These include, e.g., \brem{i)} the analysis of radial specific SFR (sSFR) profiles, obtained from \ha\ and \ewha\ determinations \citep[e.g.][]{CatalanTorrecilla17,Belfiore18}, age-dating of stellar populations via \brem{ii)} broad-band colors \citep[e.g.,][]{PelBal96,deJong96,PeletierdeGrijs98,MunozMateos07} or \brem{iii)} Lick indices \citep[e.g.][]{ThoDav06,Mor16}, and \brem{iv)} full spectral synthesis of IFS data \citep[e.g.][]{Gon14b,Gon16,SB14,SB16}.
These studies consistently report a radial decrease of sSFR in the central parts of galaxies, and a trend for an inversion (though a large scatter) of color profiles from positive (negative) in low (high) mass bulges \citep{BalPel93,BalPel94}, in qualitative agreement with the presence of negative \tlum\ 
gradients in high-mass galaxies and vice versa \citep[e.g.][]{Gon14b,Gon16}.

On the other hand, all these approaches are tied to underlying assumptions and subject to limitations. For instance, the \ewha\ is a proxy to sSFR only as long as the extinction-corrected \ha\ luminosity is a reliable SFR tracer. This requires quite specific
assumptions (continuous SF at a constant SFR since $\sim$100 Myr and no \lyc\ photon leakage, and, in the case of IFS, that 
all H$\alpha$ emission excited by stars is registered). A secondary issue is that negative stellar metallicity gradients in the bulge can readily lead to an outwardly increasing specific LyC production rate by a factor of $\sim$2, thereby mimicking positive \ha-based sSFR gradients. A further potential caveat of \brem{i} is that line-of-sight dilution by the triaxial stellar background of the bulge can naturally produce positive \ewha\ (thus also sSFR) gradients \citep[cf.][for an analogy to local blue compact dwarf galaxies]{P02}, further adding to the previous effect. 
Regarding \brem{ii}, extinction-corrected color gradients are not convertible into age gradients without assumptions on the 
star formation- and chemical enrichment history, and can strongly be affected by nebular emission \citep{Huchra77,Krueger95,P98,SdB09}. 
As for \brem{iii}, Lick indices are per se luminosity-weighted, thus primarily reflect the radial distribution of young stars. Since they are designed for instantaneously formed stellar populations (e.g., globular clusters) and not systems with prolonged SF, and potentially suffer from emission-line infilling, they appear to be of questionable applicability to LTG bulges. Finally, \brem{iv} is subject to the notorious age-metallicity-extinction degeneracy.

A further concern is that the systematization and inter-comparison of results from studies employing the methods above is partly hindered by the fact that age gradients are mostly inferred within the effective radius \reff\ instead of within 
the bulge radius \rbulge. As recently pointed out by BP18, given that \reff\ shrinks with increasing bulge-to-total ratio and is functionally coupled to the S\'ersic index \citep{Trujillo01}, the usual normalization to it can erase potentially important physical trends or lead to artificial correlations. Likewise, some studies adopt radius normalizations that might optimally serve specific science goals but are less suited to the study of the bulge component. 
As an example, \citet{Gon16} infer gradients normalized within a50, the semi-major axis of the elliptical aperture containing 50\% of the total light at 5635 \AA. Since a50 is unrelated to \rbulge\ and also cannot be converted into a photometric radius \rr\ without knowledge of the ellipticity, a transformation of age gradients from this study into \gradtL\ is not possible. 
 
The goal of our study is to infer stellar age gradients within a homogeneously defined \rbulge\ for a representative sample of local LTGs, as a step toward the systematization of the physical properties of bulges and the exploration 
of the mechanisms driving their \ioSFQ. 
In Sect. \ref{DataMethods} we briefly present our sample and analysis methodology, and in Sect.~\ref{discussion} we discuss trends between bulge age gradients and galaxy mass, and their possible interpretation in the context of \ioSFQ. The main results from this study are summarized in Sect.~\ref{summary}.
Throughout we adopt distance estimates from NASA/IPAC Extragalactic Database for H$_0$=67.8 km\,s$^{-1}$\,Mpc$^{-1}$.
\section{Analysis and main results \label{DataMethods}}
The sample used for this analysis has been studied in detail by BP18 and comprises 135 non-interacting, 
nearly face-on ($<$40$^{\circ}$) local ($\leq$130 Mpc) LTGs from the CALIFA IFS survey \citep{Sanchez12-DR1,Sanchez16-DR3} conducted with the Potsdam Multi-Aperture Spectrometer \citep[PMAS;][]{Roth05,Kelz06}.
It spans 2.6 dex in total stellar mass (8.9 $\leq$ log\mstotal\ $\leq$ 11.5) and 3~dex in bulge mass (8.3 $\leq$ log\mbstar\ $\leq$ 11.3), therefore can be considered representative of the LTG population in the local universe.

The reader is referred to BP18 for details on the photometric and spectral modeling analysis.
Here, we only recall that the isophotal radius \rbulge\ of the bulge component in our sample ranges between 
2\farcs5 and 11\farcs2 (cf. Fig. A.2 in BP18), with only four galaxies (3\%) having a bulge diameter $<$ 6\arcsec. 
Therefore, convolution with the point spread function (FWHM $\approx$ 2\farcs6, in the case of CALIFA IFS data)
does not appreciably affect our study. 
\rbulge\ was determined with the code {\sc iFit} \citep[][cf. BP18 for details]{Breda19-iFit} as the radius 
of the S\'ersic model to the bulge at an extinction-corrected surface brightness 24 $r$ \sbb. An alternative approach might have been to define \rbulge\
at the radius where the surface brightness of the bulge equals that of the disk \citep{SB14}, this would have required, however, image decomposition, which can be prone to methodological uncertainties discussed in BP18.
\begin{figure}[t]
\includegraphics[width=1.0\linewidth]{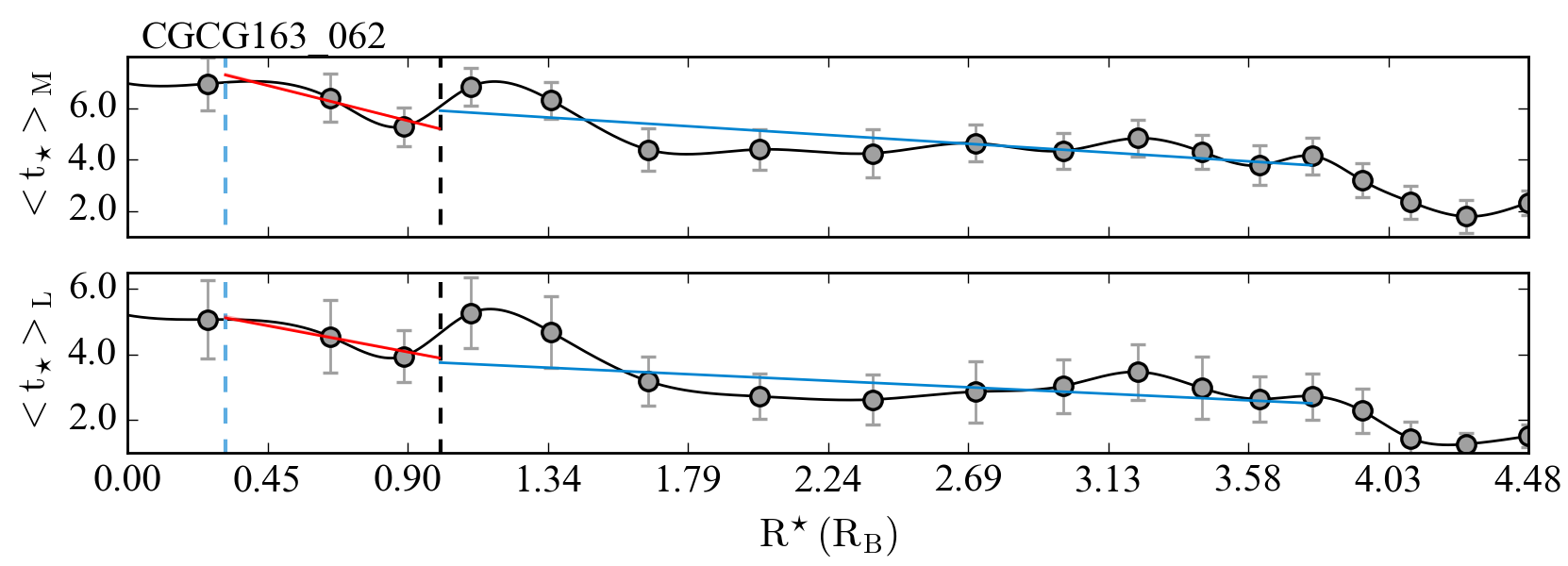}
\caption{Illustration of the derivation of \gradtM\ and \gradtL\ in the bulge through linear fits (red lines) 
between $\mathrm{R}_{\rm PSF}$ and \rb\ (blue and black vertical dashed lines, respectively) to 
spline-interpolated stellar age determinations (gray curves) within individual \isan\ (markers). 
Error bars show the standard deviation about the mean of single-spaxel determinations within each \isan.} \label{radial_profile}
\end{figure}

The spectral modeling of low-resolution ($R\!\sim\!850$) CALIFA IFS data in the V500 setup was carried out spaxel-by-spaxel with the pipeline Porto3D \citep{P13,G16a}, which invokes the population synthesis code \SL\ \citep{Cid05}. The simple stellar population libraries used consisted of templates from \citet{BruCha03} for 38 ages between 1 Myr and 13 Gyr and allowing for a time resolution of 1 Gyr for ages above 7 Gyr. As pointed out in BP18, the intrinsic $V$-band extinction in the bulge, as inferred with \SL\ for a foreground screen model was found to be relatively low (0.3 $\pm$ 0.18 mag) and not show a clear trend with \mbstar.

Single-spaxel (\sisp) determinations of the mass- and light-weighted age 
(\tmass\ and \tlum, respectively) were converted into radial profiles using an adaptation of the isophotal annuli (\isan) surface photometry technique by \citet{P02}. 
The key feature of this method lies in the computation of statistics within logarithmically equidistant isophotal zones, each corresponding to a photometric radius \rr\ (\arcsec), being defined on a reference image of the emission-line-free pseudo-continuum at 6390--6490 \AA.

Following BP18, the mean age and its uncertainty $\sigma_{\rm isan}$ within each \isan\ were determined, respectively, as the arithmetic average and standard deviation about the mean of the individual \sisp\ determinations. One advantage of this approach is that it prevents the highest-luminosity (or highest-$\Sigma_{\star}$) spaxels from dictating the result, since all spaxels subtended within an \isan\ are given equal weight. Typically, three to seven \isan\ are included within \rb.
As a next step, the mean \stmass\ and \stlight\ were spline-interpolated to a finer radius step, and a linear regression (both non-weighted and weighted by $\sigma_{\rm isan}$) was computed within $\mathrm{R}_{\rm PSF}$ $<$ \rr\ $\leq$ \rbulge, with $\mathrm{R}_{\rm PSF}$ set to 2\farcs6 in order to exclude the innermost PSF-affected part of age profiles from fits. From the latter, the mass- and light-weighted stellar age gradient (\gradtM\ and \gradtL, respectively) within the bulge was determined. Figure~\ref{radial_profile} illustrates this procedure on the example of the LTG \object{CGCG 0163-062}.

\begin{figure}[h!] 
\includegraphics[width=0.97\linewidth]{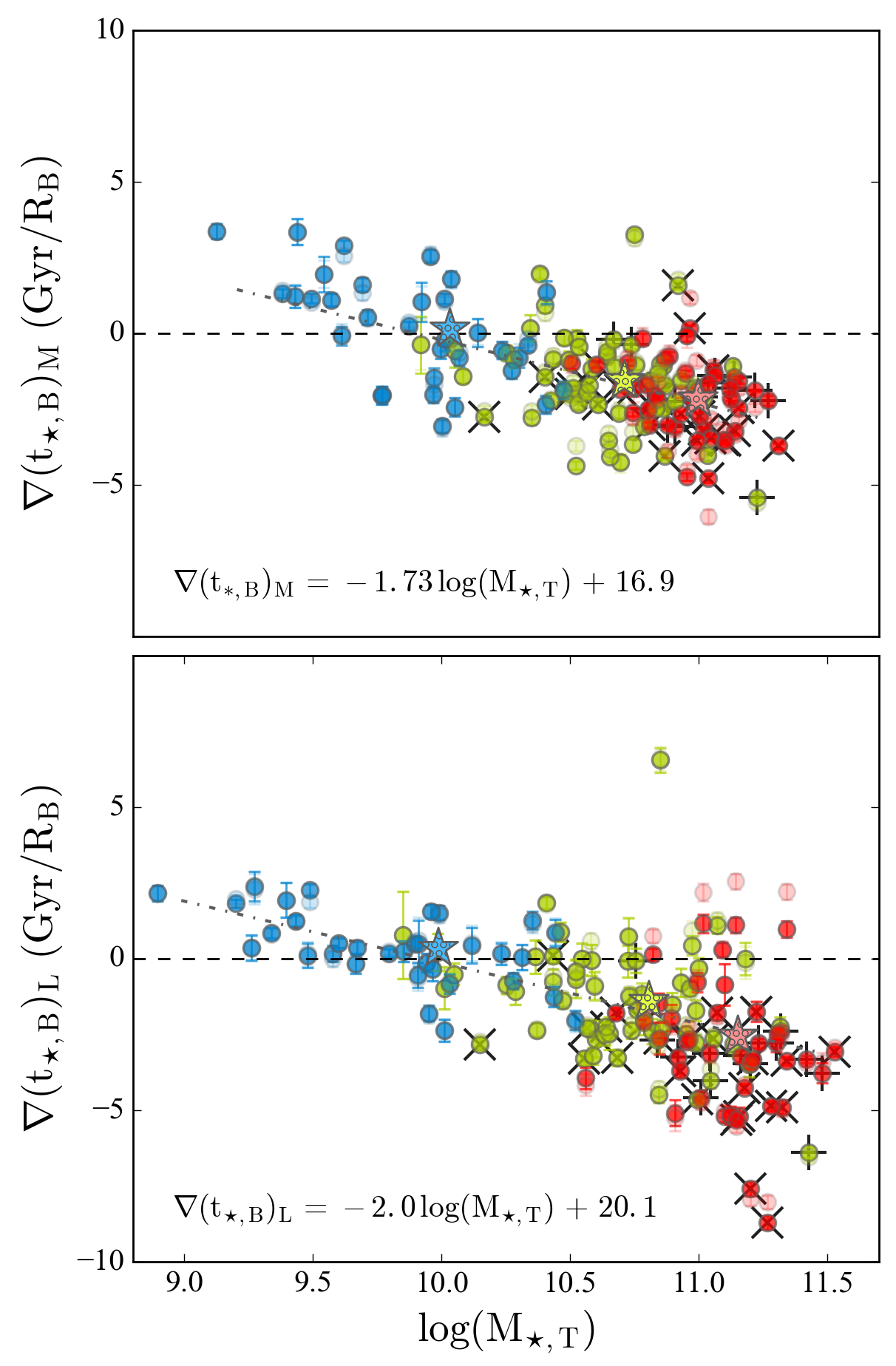}
\caption{Logarithm of total stellar mass \mstotal\ (\msun) vs. radial age gradient within the bulge radius \rbulge\ in Gyr/\rbulge, as obtained from mass- and light-weighted stellar age determinations (\gradtM\ and \gradtL; upper and lower panel, respectively). 
Symbols are color-coded according to the bulge classification scheme by BP18 (blue, green and red for \brem{iA}, \brem{iB} and \brem{iC} bulges, respectively). Semi-transparent and solid markers correspond, respectively, to determinations from non-weighted and weighted linear fits. Bulges classified by BP18 as LINER (+) and Seyfert (x) are indicated. 
Dashed-dotted lines show non-weighted linear fits to weighted \gradtM\ and \gradtL\ determinations, the arithmetic average of which for the three bulge classes is depicted by stars.}
\label{grad}
\end{figure}

Figure~\ref{grad} shows \gradtM\ and \gradtL\ vs. LTG mass log(\mstotal), with error bars depicting formal uncertainties 
from weighted and non-weighted linear fits (semi-transparent and solid markers, respectively).
The color coding corresponds to the tentative subdivision of bulges into the intervals \brem{iA}, \brem{iB} and \brem{iC} (blue, green and red, respectively) according to \dmb\ (mag), which was defined by BP18 as the difference $\mu_{\rm 0\,Gyr}$-$\mu_{\rm 9\,Gyr}$ between the mean $r$ band surface brightness of the  present-day stellar component and that of stars older than 9 Gyr. Thus, a \dmb$\approx$0 mag characterizes bulges that have completed their make-up earlier than 9 Gyr ago ($z\simeq 1.34$), whereas a \dmb\ of, say, --2.5 mag translates into a contribution of 90\% from stars younger than 9 Gyr. \citet{BP18} showed that \dmb\ tightly correlates with physical and evolutionary properties of LTG bulges (e.g., stellar age, surface density and mass) and proposed it as a convenient means for their classification.
The most massive, dense and old bulges fall onto the \brem{iC} class (\dmb\ $\geq$ --0.5 mag) whereas bulges classified 
as \brem{iA} (\dmb\ $\leq$ --2.5 mag) are the youngest and reside in the least massive LTGs. Spectroscopically classified after \citet{BPT}, nearly all \brem{iA} bulges fall on the locus of H{\sc ii} regions, whereas almost all bulges classified as LINER (+) and Seyfert (x) are hosted by LTGs with log(\mstotal) $>$ 10.5, as apparent from Fig.~\ref{grad}.

Linear fits to weighted determinations of age gradients (dashed-dotted lines) yield the relations 
\gradtM\ = --(1.73 $\pm$ 0.15) $\cdot$ log(\mstotal) + (17.0 $\pm$ 1.56) and \gradtL\ = --(2.0 $\pm$ 0.15) $\cdot$ log(\mstotal) + (20.1 $\pm$ 1.56).

Despite a large scatter, both panels of Fig.~\ref{grad} consistently reveal a trend for an inversion of age gradients 
from positive to negative values with increasing galaxy mass: whereas \brem{iA} bulges hosted by lower-\ms\ LTGs show in their majority flat or positive 
age gradients, the opposite is the case for \brem{iB} \& \brem{iC} bulges for which both \gradtM\ and \gradtL\ are generally negative with an average value of $\sim$--2 Gyr/\rb\ above log\mstotal\ $\simeq$ 10.5.

We note that this trend is highly unlikely to be driven by the outwardly increasing contribution of the star-forming disk, given that profile decomposition by BP18 yields that in most cases the latter provides a small fraction of the $r$-band emission at \rb, and \tmass\ determinations are relatively insensitive to young stellar populations.
Moreover, would disk contamination be strongly affecting mass-weighted age gradients within \rbulge, then one would expect low-mass (\brem{iA}) bulges to be most prone to this effect and show negative \gradtM\ values, which is the opposite of the evidence from Fig. \ref{grad}.
%
\section{Discussion \label{discussion}}
Over the past years, observational evidence has been accumulating for a dependence of the slope of stellar age gradients on galaxy mass and luminosity (cf. Sect.~\ref{intro}). The present study adds further insights into this subject by virtue of the fact that it investigates for the first time a representative sample of local LTGs and determines age gradients within a clear-cut defined bulge radius.
On the other hand, one should bear in mind some methodological limitations of this study. Of those, especially important is the age-metallicity-extinction degeneracy being inherent to all purely stellar population spectral synthesis codes (cf. BP18 for a further discussion).
In the case of \starlight, typical uncertainties of 0.1-0.15 dex on age determinations \citep{Cid05,Cid14} within individual \isan\ could propagate into errors in \gradtL\ and \gradtM\ at a level that we estimate to be $\sim$20\%.
The same applies to systematic errors arising from the neglect of nebular continuum emission: as demonstrated by \citet{CGP19}, 
already at a modest level of nebular contamination (\ewha\ $\simeq$ 40-60 \AA) the \tmass\ inferred from \starlight\ can be overestimated by 0.2-0.3 dex. Additionally, several of the here analyzed LTGs show within \rb\ significant departures from a linear increase of stellar age with radius, thus linear fits (cf. Fig. \ref{radial_profile}) yield in these cases rather coarse estimates to \gradtM\ and \gradtL.

The large scatter ($\sigma$ $\sim$1.7 Gyr/\rb) in Fig.~\ref{grad} does not permit to associate the inversion point from positive to negative age gradients with a sharp threshold in \mstotal. 
However, judging from the linear fits in Fig.~\ref{grad}, this transition occurs 
at approximately log(\mstotal/\msun) $\simeq$ 10, i.e. the empirically suggested border between SF- and AGN/LINER-dominated LTG bulges (BP18).
This is consistent with the notion that the inversion of stellar age gradients marks the epoch when accretion-powered nuclear activity starts taking over the gas excitation in bulges, then gradually leading to the extinction of their SF activity. The fact that negative age gradients are naturally arising in an \ioSFQ\ scenario reinforces this conjecture, even though this alone is no compelling evidence for a causal link between both phenomena.
In the light of the trends in Fig. \ref{grad} it appears worthwhile to extend this study with metallicity determinations both in the stellar and nebular component, which is a forthcoming task. Interestingly, \citet{Tis16} find from cosmological simulations evidence for inside-out disk formation, with simulated discs with a mass around $10^{10}$ \msun\ showing steeper negative stellar metallicity gradients, in agreement with observational results \citep[e.g.][]{Gon15}.
\begin{figure}
\includegraphics[width=0.97\linewidth]{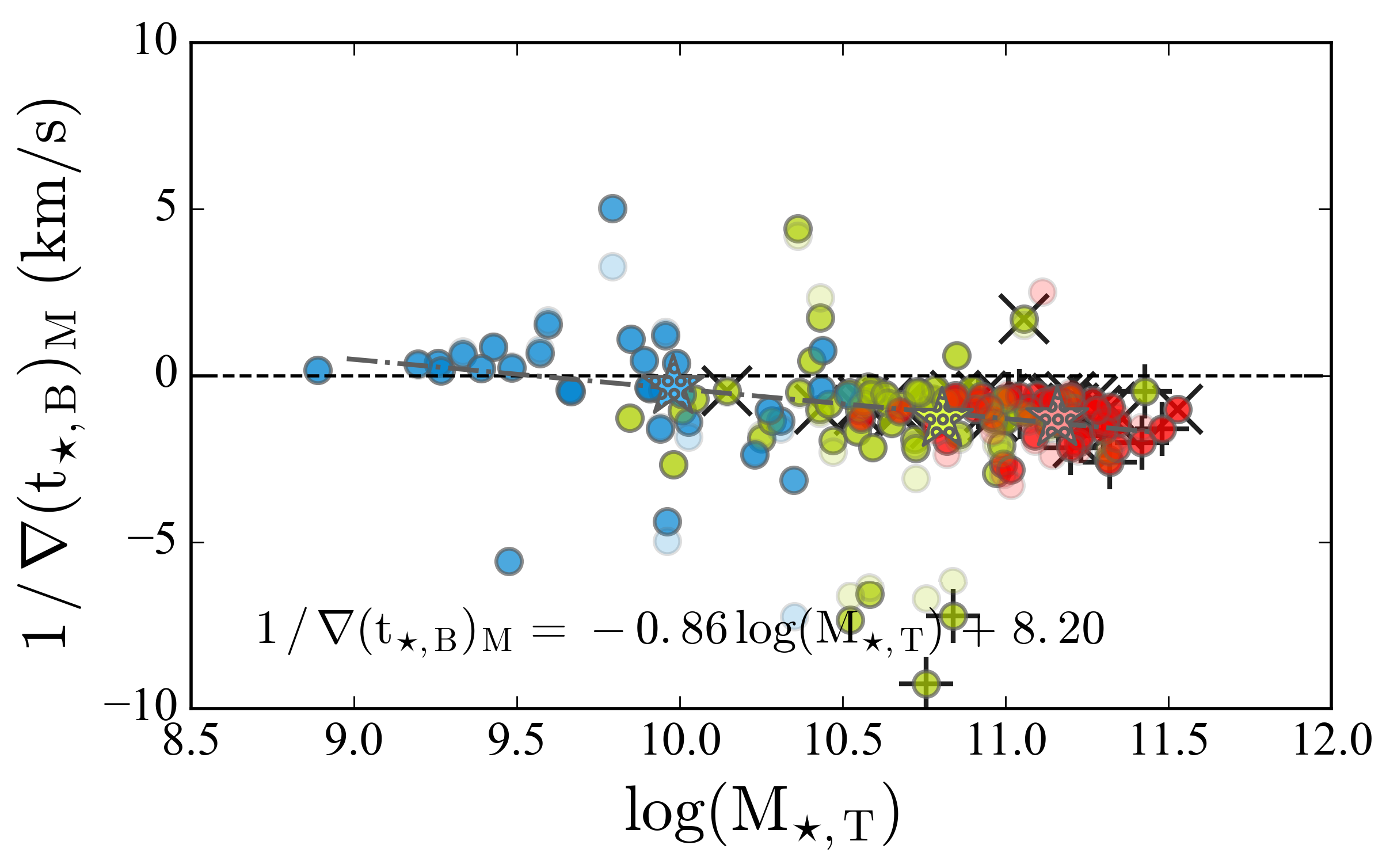}
\caption{Logarithm of total stellar mass \mstotal\ vs. age gradient (determinations from weighted linear fits) translated into an 
average inside-out SF quenching velocity (\vq; negative values) or outside-in SF shrinking velocity (positive values) in the bulge. Dashed-dotted lines show linear fits. The meaning of symbols is identical to that in Fig.~\ref{grad}.
}
\label{vq}
\end{figure}

One may consider two simple limiting cases for the bulge growth prior to the onset of \ioSFQ, the first one employing the assumption that the sSFR within \rbulge\ has initially been roughly spatially uniform (thus \gradtM\ $\approx$ 0) and the second one assuming an inside-out bulge assembly process, thus a negative \gradtM\ of genetic origin. 
In a simple gedankenexperiment, one can then convert in the first case the present-day \gradtM\ 
into a mean radial velocity \vq\ (\kmsec) for a spheric-symmetric outwardly (inwardly) propagating front of SF quenching (shrinking).

As apparent from Fig.~\ref{vq}, for the majority of bulges above log(\mbstar) $\approx$ 10 (\brem{iB} and \brem{iC} class in the notation by BP18), this \vq\ is in the range of 1-2 \kmsec, which is $\sim$2 dex lower than the central stellar velocity dispersion 
of LTG bulges \citep[150-200 \kmsec,][]{FB16}\footnote{Interestingly, in \citet{G16b} we simulated the formation 
of ETGs in a toy model assuming inside-out SF since 13.5 Gyr at a mean velocity of 2 \kmsec, which was chosen such as to reproduce the typical radius of these galaxies. The fact that this assumed velocity turns out to be fairy close to the here observationally estimated mean \vq\ for LTG bulges, if not a mere coincidence, raises the speculation of whether stellar spheroids both form and quench in an inside-out manner at a speed on the other of the sound speed in the neutral gas.}.
This prima facie points against a brief energetic event (e.g., an AGN-driven blast wave or series of starbursts) that rapidly swept gas out of the bulge, thereby leading to an abrupt cessation of SF out to \rb.
The \gradtM\ inversion point at a present-day stellar mass of log(\mstotal) $\simeq$ 10 translates by the relations log(\mbulge) = 1.22 $\cdot$ log(\mstotal) - 2.87 and \mtmass\ = 2.3 $\cdot$ log(\mbulge) - 13.44 by BP18 
to an age of $\sim$9.6 Gyr ($z \approx 0.4$)
for the onset of \ioSFQ.
This suggests that, on the statistical average, LTGs with a present-day log(\mstotal) $\geq$ 10 have 
entered their \ioSFQ\ phase no later than $\sim$4 Gyr after the Big Bang. 
To reconcile this with the observed age contrast between bulge center and periphery ($\sim$3 Gyr; cf. Fig. \ref{grad}) 
in the most massive ($> 10^{11}$ \msun) LTGs, one has then to assume that \ioSFQ\ in these systems
has started at a cosmic age of $\la$ 1 Gyr ($z$ $\ga$ 5.7). The hypothesis that this process was driven by negative AGN feedback appears to be compatible with existing data, even though the peak of the average SMBH growth rate has occurred only at $z$ $\la3$ \citep[e.g.,][]{Shankar09}. 

Admittedly, the assumption of a nearly constant \vq\ is simplistic: for instance, by analogy to a SF-driven supershell that spherically expands against an ambient medium of constant density, one would expect the radius to grow as $t^{2/5}$ and $t^{3/5}$ for the case of, respectively, an instantaneous and continuous injection of mechanical energy \citep[e.g.][see also \textcolor{myblue4}{McCray \& Kafatos 1987}]{deYoungHeckman1994}. 
Furthermore, a stable, spheric-symmetrically expanding \ioSFQ\ front is improbable in the presence of a turbulent multi-phase gas medium exposed to strongly directional outflows from an AGN, as several observations indicate \citep[e.g.,][see Kormendy \& Ho 2013 for a review]{Kehrig12}.

As for the second hypothesis, it assumes a superposition of a pre-existing negative \gradtM\ with that subsequently arising through \ioSFQ. Discriminating between both is clearly a formidable task, perhaps only possible through a combined chemodynamical decomposition of stellar populations and the nebular emission they are associated with.
Obviously, if the observed \gradtM's of $\sim$3 Gyr/\rb\ are primarily of genetic origin, then massive (\brem{iB} and \brem{iC}) bulges cannot have formed quasi-monolithically \citep[$<$1 Gyr,][for a review]{KorKen04}, but instead over a prolonged phase of 2-3 Gyr that lasted until $z$ $\approx$ 2.2, as suggested by BP18. These authors proposed that bulges arise out of galactic disks on a timescale that is inversely related to the present-day LTG mass, with massive \brem{iC} bulges forming in the most massive galaxies first, whereas the least massive \brem{iA} bulges still assembling out of gaseous and stellar material from the disk.
This minimum timescale of 2-3 Gyr for the make-up of bulges is probably also consistent with the two-phase galaxy formation scenario by 
\citet{Oser10}, if the initial ($z\ga$2) dissipational phase of in-situ SF in their simulations is chronologically associated with the dominant phase of bulge formation in the most massive LTGs. The same applies to the timescales predicted by bulge formation scenarios envisaging inward migration and coalescence of massive ($\sim 10^9$ \msun) SF clumps emerging out of violent disk instabilities \citep{Bournaud07,Mandelker14}. As pointed out in BP18, a natural consequence from inward migration of SF clumps from the disk are negative age gradients in massive \brem{iC} bulges, in agreement with the evidence from Fig.~\ref{grad}.
%
%

The large scatter in \gradtM\ in Fig.~\ref{grad}), even though partly due to observational and methodological uncertainties, suggests that there are multiple evolutionary pathways from bulges bracketed by the two limiting cases of slow and quick \ioSFQ. The spread of determinations at the l.h.s. of the diagram (\gradtM\ $\ga$ 0) might reflect a various degree of intensity and confinement of SF to the center of low-mass bulges, possibly regulated by 
the inflow rate of intra-clump gas from the disk \citep{Hopkins2012} and influenced by minor wet mergers with gaseous entities from the galaxy itself and its surroundings.\newline 
As for bulges with \gradtM\ $<$ 0, the documented spread in \gradtM\ might be accounted for by a superposition of different non-mutually exclusive processes. These include both negative and positive AGN feedback, the latter in the sense of fast nuclear outflows plunging into the ambient dense gas and triggering localized SF in the periphery of the bulge, this way promoting its growth \citep{Maiolino17} and further steepening negative age gradients.
The role of bars for this phase and over the whole bulge evolution is most certainly another important unknown: \citet{GA01} found mostly negative $U$-$B$ and $B$-$V$ color gradients, with flat or positive gradients mostly in barred galaxies.

The expected imprints of morphological quenching or gas starvation (cf. Sect. \ref{intro})
on stellar age gradients in the bulge has, to our best knowledge, not been quantitatively explored from the theoretical point of view. Whereas both of these processes act towards shutting off SF all over \rbulge, thus probably have no dominant effect on pre-existing age gradients, 
it cannot be excluded that they favor, over a certain period at least, a non-uniform sSFR within the bulge, eventually leaving an imprint on \gradtM\ and contributing to its scatter. The resolution of current cosmological simulations is insufficient for spatially resolving the bulge 
and carrying out a comparative investigation of the expected \gradtt\ for the above scenarios. 
Moreover, contrary to an AGN, one cannot switch on/off gas inflow onto the bulge in order to quantitatively assess the effect of inhibition 
of gas transfer from the cosmic web \citep{Dekel09} or the local galaxy environment \citep[e.g.,][]{Peng15}. 
Additionally, the investment in computational time for higher-resolution simulations appears justifiable only after an 
improvement of contemporary models, which in some respects suffer from significant deficits.
For instance, simulations on larger spatial scales (out to \reff) by \citet{Hirschmann13} and \citet{Choi17} always predict strongly positive age gradients in massive (log(\mstotal/\msun)$>$10) galaxies \citep{Hirschmann15}, which indicates that better prescriptions or additional physical processes need to be implemented.
\section{Summary and conclusions \label{summary}}
The trend for an inversion of the slope of radial stellar age gradients from positive values in young low-mass bulges to negative values for old massive bulges, here solidified through a homogeneous analysis of representative sample of local late-type galaxies from the CALIFA survey, likely encodes crucial information on the physical drivers of bulge growth and the ensuing inside-out cessation of star-forming activity.

For massive bulges (log\mstotal$>$10), we distinguish between two simple limiting cases, the first one attributing  
stellar age gradients to an inside-out quenching of star formation, and the second one viewing them as relics from the bulge formation process.
In the first case, and assuming that prior to \ioSFQ\ bulges had spatially uniform age, the observed gradients translate into a mean inside-out \SFQ\ velocity of 1--2 \kmsec, which is consistent with a gradual evacuation or thermalization a cold gas being susceptible to star formation. If due to negative AGN feedback, this could hint at a mild growth of super-massive black holes in tandem with the bulge. However, alternative mechanisms for the shut-off of star formation (e.g., morphological quenching or inhibition of cold gas inflow from the halo and the cosmic web) cannot be excluded given the lack of quantitative theoretical constraints on the radial age patterns these processes could leave behind.

In the second scenario, i.e. assuming that the observed age gradients portray the bulge formation process, then the 
range of values inferred ($\sim$3 Gyr/\rb) argue against a quasi-monolithic build-up and point instead to a prolonged 
formation phase that could be driven by a superposition of inwards migration of star-forming clumps from the disk 
and in situ star formation. Quite importantly, if the observed age gradients in bulges are mainly of genetic origin, then 
one cannot rule out a quick \ioSFQ\ episode following an energetic AGN outburst, given that such a process would leave virtually no imprints on pre-existing age patterns in the stellar component.

As the large scatter (1.7 Gyr/\rb) of age gradients at a given mass suggests, the evolutionary pathways of galaxy bulges are far more complex, and possibly shaped by a mixture of the two aforementioned limiting 
scenarios with, for instance, positive AGN feedback, bulge-bar interaction and possibly also morphological quenching and gas starvation. A further exploration of the age gradient vs. stellar mass relation with higher-resolution IFS data and detailed numerical simulations, incorporating realistic recipes for star formation and AGN-driven feedback appears to be of considerable interest and fundamental to the development of a coherent picture on the assembly history of late-type galaxies and their structural components.

\begin{acknowledgements}
We thank the European taxpayer, who in the spirit of solidarity between EU countries is offering to Portugal a substantial fraction of the financial resources that allowed it to sustain a research infrastructure in astrophysics. Specifically, the major part of this work was carried out at an institute whose funding is provided to $\sim$85\% by the EU via the FCT (Funda\c{c}\~{a}o para a Ci\^{e}ncia e a Tecnologia) apparatus, through European and national funding via FEDER through COMPETE by the grants UID/FIS/04434/2013 \& POCI-01-0145-FEDER-007672 and PTDC/FIS-AST/3214/2012 \& FCOMP-01-0124-FEDER-029170. Additionally, this work was supported by FCT/MCTES through national funds (PIDDAC) by this grant UID/FIS/04434/2019.
We are grateful to Dr. Dimitri Gadotti, Prof. Daniel Schaerer and Dr. Andrew Humphrey for valuable comments.
We acknowledge supported by European Community Programme (FP7/2007-2013) under grant agreement No. PIRSES-GA-2013-612701 (SELGIFS). 
I.B. was supported by the FCT PhD::SPACE Doctoral Network (PD/00040/2012) through the fellowship PD/BD/52707/2014 
funded by FCT (Portugal) and POPH/FSE (EC) and by the fellowship CAUP-07/2014-BI in the context of the FCT project PTDC/FIS-AST/3214/2012 \& FCOMP-01-0124-FEDER-029170. 
P.P. was supported through Investigador FCT contract IF/01220/2013/CP1191/CT0002 and by a contract 
that is supported by FCT/MCTES through national funds (PIDDAC) and by grant PTDC/FIS-AST/29245/2017.
SA gratefully acknowledge support from the Science and Technology Foundation (FCT, Portugal) through the research grants PTDC/FIS-AST/29245/2017 and UID/FIS/04434/2019.
P.L. acknowledges support by DL57/2016/CP1364/CT0010.
This study uses data provided by the Calar Alto Legacy Integral Field Area (CALIFA) survey (califa.caha.es), 
funded by the Spanish Ministry of Science under grant ICTS-2009-10, and the Centro Astron\'omico Hispano-Alem\'an.
It is based on observations collected at the Centro Astron\'omico Hispano Alem\'an (CAHA) at Calar Alto, operated jointly 
by the Max-Planck-Institut f\"ur Astronomie and the Instituto de Astrof\'isica de Andaluc\'ia (CSIC).
This research has made use of the NASA/IPAC Extragalactic Database (NED) which is operated by the Jet Propulsion Laboratory, 
California Institute of Technology, under contract with the National Aeronautics and Space Administration.
\end{acknowledgements}

\end{document}